\documentclass[%
 reprint,
 amsmath,amssymb,
 aps,
 prl
 longbibliography,
]{revtex4-2}

\usepackage[dvipdfmx,hiresbb]{graphicx}
\usepackage{dcolumn}
\usepackage{bm}
\usepackage[T1]{fontenc}
\usepackage[dvipdfmx]{hyperref}
\hypersetup{
colorlinks = true,
linkcolor = blue,
citecolor = blue,
}

\begin{document}


\title{Reply to ``Comment on `Theoretical analysis of quantum turbulence using the Onsager ideal turbulence theory' ''}

\author{Tomohiro Tanogami}
\affiliation{%
Department of Physics, Kyoto University, Kyoto 606-8502, Japan
}%




\date{\today}
\begin{abstract}
We refute the criticism expressed in a Comment by Krstulovic, L'vov, and Nazarenko [arXiv:2107.10598] on our paper [Phys.\ Rev.\ E \textbf{103}, 023106 (2021)].
We first show that quantization of circulation is not ignored in our analysis.
Then, we propose a more sophisticated analysis to avoid a subtle problem with the regularity of the velocity field.
We thus defend the main results of our paper, which predicts the double-cascade scenario where the \textit{quantum stress cascade} follows the Richardson cascade.
We also provide a conjecture on the relation between the Kelvin-wave cascade and the quantum stress cascade.
\end{abstract}

\pacs{Valid PACS appear here}

\maketitle

In our recent paper \cite{tanogami2021theoretical}, we investigated the quantum turbulence described by the Gross-Pitaevskii equation (GPE) using a phenomenological approach based on the Onsager ``ideal turbulence'' theory \cite{Onsager_1949,Eyink_Sreenivasan,Eyink_Review}.
Krstulovic \textit{et al.}\ in their Comment \cite{krstulovic2021comment} claim that there are some physical problems with the double-cascade scenario proposed in \cite{tanogami2021theoretical}.
Their arguments are summarized as follows:
(i) the dynamics of the quantum (dispersive) Euler equations is mathematically equivalent to that of the GPE only in the absence of vortices;
(ii) the assumption of the Besov regularity of the velocity field does not hold in the presence of quantum vortices;
(iii) the existence of the \textit{quantum stress cascade} is questionable because the same analysis can be applied to the linear Schr\"odinger equation;
and (iv) if the quantum stress cascade displaying a $k^{-3}$ spectrum exists, it is overwhelmed by the Kelvin-wave cascade, which displays a shallower spectrum.
In this Reply, we show that (i) and (iii) are not justified.
For (ii), which is a subtle point, we propose a more sophisticated analysis using the density-weighted velocity $\sqrt{\rho}{\bf v}$.
While we agree on point (iv), it is already mentioned in \cite{tanogami2021theoretical}.
In this regard, we provide a conjecture on the relation to the Kelvin-wave cascade.

\paragraph*{Reply to (i).}
Krstulovic \textit{et al.}\ claim that the analysis of \cite{tanogami2021theoretical} ignores quantum vortices because \textit{the dynamics of the dispersive Euler equation is mathematically equivalent to one of the GP equation only in absence of vortices}.
However, this statement is not accurate because the solutions of the quantum Euler equations contain the solutions of the GPE as a proper subset.
In other words, while the solutions of the quantum Euler equations are not generally equivalent to those of the GPE because arbitrary circulation is allowed in the quantum Euler equations, the former that satisfy the quantization condition are equivalent to the latter.

To see this, let $\{\rho_\Psi(\cdot,t),{\bf v}_\Psi(\cdot,t)\}$ be the superfluid mass density and velocity fields at time $t$ obtained from the condensate's complex wave function $\Psi(\cdot,t)$ that satisfies the GPE via the Madelung transformation
\begin{equation}
\Psi({\bf x},t)=\sqrt{\dfrac{\rho_\Psi({\bf x},t)}{m}}\exp(i\theta({\bf x},t))
\end{equation}
with ${\bf v}_\Psi=(\hbar/m)\nabla\theta$, where $m$ denotes the boson mass.
Note that $\{\rho_\Psi(\cdot,t),{\bf v}_\Psi(\cdot,t)\}$ satisfies the following properties:
(i) on the nodal lines where superfluid is absent, i.e., $\{{\bf x}|\rho_\Psi({\bf x},t)=0\}$, the superfluid velocity ${\bf v}_\Psi$ is \textit{obviously} not defined, and (ii) for any closed loop $C$ that does not pass through the nodal lines, the circulation $\Gamma_C:=\oint_C{\bf v}_\Psi({\bf r},t)\cdot d{\bf r}$ is an integer multiple of $2\pi\hbar/m$ because of the single-valuedness of $\Psi$.
Conversely, the single-valued function $\Psi(\cdot,t)$ is uniquely recovered from $\{\rho_\Psi(\cdot,t),{\bf v}_\Psi(\cdot,t)\}$ up to an overall phase factor because of the quantization condition for $\Gamma_C$ \cite{wallstrom1994inequivalence}.
These properties hold at any time $t$, so there is one-to-one correspondence between $\Psi$ and $\{\rho_\Psi,{\bf v}_\Psi\}$ up to an overall phase factor.
In general, a function $\tilde{\Psi}$ constructed from the solutions of the quantum Euler equations $\{\rho,{\bf v}\}$ becomes multivalued because ${\bf v}$ does not necessarily satisfy the quantization condition.
Thus, the solutions of the quantum Euler equations contain the solutions of the GPE as a proper subset \cite{wallstrom1994inequivalence}.
This means that the quantum Euler equations can describe the motion of the vortex lines.
We remark that, as long as we consider $\{\rho_\Psi,{\bf v}_\Psi\}$, we do not have to add by hand ``extremely complex moving boundary conditions'' \cite{krstulovic2021comment} to describe the quantum vortex dynamics because $\{\rho_\Psi,{\bf v}_\Psi\}$ automatically satisfies the quantization condition.

Related to this point, the statement in \cite{tanogami2021theoretical} that \textit{We investigate three-dimensional quantum turbulence as described by the Gross-Pitaevskii model} is not misleading.
This statement means that $\{\rho,{\bf v}\}$ in \cite{tanogami2021theoretical} represents $\{\rho_\Psi,{\bf v}_\Psi\}$ in the notation used here.
More precisely, while the analysis of \cite{tanogami2021theoretical} itself is applicable even to the solutions that do not satisfy the quantization condition (see also \cite{tanogami2021van}), only the interpretation of the results for $\{\rho_\Psi,{\bf v}_\Psi\}$ is presented in \cite{tanogami2021theoretical}.
It is also not justified to claim that the results of \cite{tanogami2021theoretical} are not applicable at scales smaller than $\ell_i$.
At least the main results that do not rely on the Besov regularity condition of the velocity field, such as the ``Kolmogorov's $4/5$ law'' for quantum turbulence (Eq. (28) of \cite{tanogami2021theoretical}), are completely valid even at scales smaller than $\ell_i$.

We finally remark that although Krstulovic \textit{et al.}\ cite \cite{wallstrom1994inequivalence} to support their argument, \cite{wallstrom1994inequivalence} explicitly states that the solutions of the Madelung hydrodynamic equations contain the solutions of the Schr\"odinger equation as a proper subset.

\paragraph*{Reply to (ii).}
Krstulovic \textit{et al.}\ question the validity of the assumption of the H\"older continuity of the velocity field, which is described in the main text of \cite{tanogami2021theoretical}.
While they may be right on that point, the use of the H\"older continuity is only for clarity of presentation.
The actual analysis is done using Besov regularity in Appendix B of \cite{tanogami2021theoretical}.
While Krstulovic \textit{et al.}\ claim that the assumption of Besov regularity is also far from being justified physically, the condition is expected to hold up to $p=6$ at least in the scale range $\ell_i\ll\ell\ll\ell_{\mathrm{large}}$ according to recent experiments \cite{Rusaouen_etal_2017}.
We agree that the Besov regularity is still not verified in the scale range $\ell_{\mathrm{small}}\ll\ell\ll\ell_i$.
To avoid this subtle problem, we propose a more sophisticated analysis using the density-weighted velocity ${\bf w}:=\sqrt{\rho}{\bf v}$.
Instead of assuming the Besov regularity of the velocity field ${\bf v}$, we here assume the Besov regularity of the density-weighted velocity ${\bf w}$:
\begin{align}
\|\delta{\bf w}({\bf r};\cdot)\|_p\sim A_pw_0\left(\dfrac{|{\bf r}|}{L}\right)^{\sigma_p}\quad\text{as}\quad |{\bf r}|/L\rightarrow0
\end{align}
with a dimensionless constant $A_p$ for $p\in[1,\infty]$ and $\sigma_p\in(0,1]$.
Here, $\|\cdot\|_p=\langle|\cdot|^p\rangle^{1/p}$, where $\langle\cdot\rangle$ denotes the spatial average, $\delta{\bf w}({\bf r};{\bf x}):={\bf w}({\bf x}+{\bf r})-{\bf w}({\bf x})$, and $w_0$ denotes a typical density-weighted velocity.
Note that the validity of this condition has been partially verified in several numerical simulations \cite{Kobayashi-Ueda,krstulovic2016grid,Leoni,Shukla_etal_2019}.
Even for this case, we can predict the double-cascade scenario, where the quantum stress cascade follows the Richardson cascade, as shown in Appendix.

\paragraph*{Reply to (iii).}
Krstulovic \textit{et al.}\ question the possibility of the quantum stress cascade by pointing out that the same analysis for the quantum stress cascade as in \cite{tanogami2021theoretical} could be applied to the linear Schr\"odinger equation because the quantum stress is independent of the GP nonlinearity.
Note that their argument also holds even for the Richardson cascade, induced by the momentum flux $\rho{\bf v}{\bf v}$, because the nonlinear term $\rho{\bf v}{\bf v}$ of the quantum Euler equations is also independent of the GP nonlinearity.
Because the existence of the Richardson cascade has been well confirmed in many experiments and simulations, their argument is not justified.
The fact that a nonlinear term becomes linear by a nonlinear transformation does not mean that there is no cascade induced by that nonlinear term.

\paragraph*{Reply to (iv).}
We agree that the $k^{-3}$ spectrum for the quantum stress cascade can become shallower because of the Kelvin-wave oscillations.
The possibility of such \textit{depletion of nonlinearity} has already been mentioned in \cite{tanogami2021theoretical}.
We also note that \cite{tanogami2021theoretical} does not deny the existence of the Kelvin-wave cascade.
It only states the possibility that the quantum stress cascade is related to the Kelvin-wave cascade.

We now remark on the relation between the Kelvin-wave cascade and the quantum stress cascade.
Note that if the Kelvin-wave cascade exists, there must be a corresponding scale-to-scale energy flux other than the \textit{deformation work} $\Pi_\ell$, which induces the Richardson cascade.
In the large-scale kinetic energy budget equation Eq.\ (36) of \cite{tanogami2021theoretical}, the only energy flux specific to the quantum case is the \textit{quantum baropycnal work} $\Lambda^{(\Sigma)}_\ell$, which induces the quantum stress cascade.
From this fact, we expect that the quantum stress cascade corresponds to the Kelvin-wave cascade.
More precisely, because the theoretical analysis of the Kelvin-wave cascade has been developed for the vortex filament model \cite{KS,LN}, where compressible effects are ignored, we conjecture that the incompressible part of the quantum baropycnal work $\Lambda^{(\Sigma)}_\ell$ corresponds to the energy flux of the Kelvin-wave cascade.

\paragraph*{Conclusion.}
In summary, we have shown that:
\begin{itemize}
	\item[(i)] The solutions of the quantum Euler equations contain the solutions of the GPE as a proper subset.
	In other words, the quantum Euler equations can describe the motion of the vortex lines, and thus quantization of circulation is not ignored in our paper \cite{tanogami2021theoretical}.
	\item[(ii)] The Besov regularity of the velocity field is indeed still not verified. 
	To avoid this subtle problem, we can develop a more sophisticated analysis using the density-weighted velocity $\sqrt{\rho}{\bf v}$ and obtain the same result as in \cite{tanogami2021theoretical}.
	\item[(iii)] Krstulovic \textit{et al.}'s criticism of the quantum stress cascade based on the fact that the same analysis can be applied to the linear Schr\"odinger equation is not justified.
	\item[(iv)] The possibility that the $k^{-3}$ spectrum for the quantum stress cascade becomes shallower because of the Kelvin-wave oscillations has already been mentioned in \cite{tanogami2021theoretical}.
	In this regard, the incompressible part of the quantum baropycnal work $\Lambda^{(\Sigma)}_\ell$ is expected to correspond to the energy flux of the Kelvin-wave cascade.
\end{itemize}
We thus clarify the validity of the main results of our paper, which predicts the double-cascade scenario where the quantum stress cascade follows the Richardson cascade.
We hope that experiments and numerical simulations will be conducted to verify our predictions.

\begin{acknowledgements}
The author thanks Michikazu Kobayashi, Makoto Tsubota, Kazuya Fujimoto, Gregory Eyink, and Shin-ichi Sasa for useful discussions.
The present study was supported by JSPS KAKENHI Grant No. 20J20079, a Grant-in-Aid for JSPS Fellows.
\end{acknowledgements}

\appendix
\section{Analysis using the density-weighted velocity}
Here, we provide the details of the more sophisticated analysis using the density-weighted velocity.
We consider the quantum Euler equations obtained after the Madelung transformation of the GPE (the notation is the same as in \cite{tanogami2021theoretical}):
\begin{equation}
\partial_t\rho+\nabla\cdot(\rho{\bf v})=0,
\end{equation}
\begin{equation}
\partial_t(\rho{\bf v})+\nabla\cdot(\rho{\bf v}{\bf v})=-\nabla p+\nabla\cdot{\bf \Sigma}+{\bf f}.
\end{equation}
Let ${\bf w}:=\sqrt{\rho}{\bf v}$ be the density-weighted velocity.
The quantum Euler equations can be rewritten in terms of this quantity as
\begin{equation}
\partial_t\rho+\nabla\cdot(\sqrt{\rho}{\bf w})=0,
\label{qEuler-J-1}
\end{equation}
\begin{equation}
\partial_t(\sqrt{\rho}{\bf w})+\nabla\cdot({\bf w}{\bf w})=-\nabla p+\nabla\cdot{\bf \Sigma}+{\bf f}.
\label{qEuler-J-2}
\end{equation}

To investigate the energy transfer across scales, we take a coarse-graining approach that can resolve turbulent fields both in space and in scale.
For any field ${\bf a}({\bf x})$, we define a coarse-grained field at length scale $\ell$ as
\begin{equation}
\bar{{\bf a}}_\ell({\bf x}):=\int_{\Omega}d^3{\bf r}G_\ell({\bf r}){\bf a}({\bf x}+{\bf r}),
\label{coarse-graining}
\end{equation}
where $G:\Omega\rightarrow [0,\infty)$ is a smooth symmetric function supported in the open unit ball with $\int_{\Omega}G=1$, and $G_\ell({\bf r}):=\ell^{-3}G({\bf r}/\ell)$ is the rescaling defined for each $\ell>0$.
Coarse-graining of (\ref{qEuler-J-1}) and (\ref{qEuler-J-2}) gives
\begin{equation}
\partial_t\bar{\rho}_\ell+\nabla\cdot\overline{(\sqrt{\rho}{\bf w})}_\ell=0,
\label{qEuler-J-1-coarse-grained}
\end{equation}
\begin{equation}
\partial_t\overline{(\sqrt{\rho}{\bf w})}_\ell+\nabla\cdot\overline{({\bf w}{\bf w})}_\ell=-\nabla \bar{p}_\ell+\nabla\cdot\bar{{\bf \Sigma}}_\ell+\bar{{\bf f}}_\ell.
\label{qEuler-J-2-coarse-grained}
\end{equation}
We introduce the following density-weighted coarse-grained variable $\widehat{\bf w}_\ell$ to obtain a simple physical interpretation:
\begin{equation}
\widehat{\bf w}_\ell:=\dfrac{\overline{(\sqrt{\rho}{\bf w})_\ell}}{\sqrt{\bar{\rho}_\ell}}.
\end{equation}
Note that $\widehat{\bf w}_\ell$ is different from the density-weighted average $\overline{(\sqrt{\rho}{\bf w})_\ell}/\overline{\sqrt{\rho}}_\ell$.
Because $\sqrt{\rho}$ is a concave function of $\rho$, we find that $\overline{\sqrt{\rho}}_\ell\le\sqrt{\bar{\rho}_\ell}$.
We can rewrite (\ref{qEuler-J-1-coarse-grained}) and (\ref{qEuler-J-2-coarse-grained}) in terms of $\widehat{\bf w}_\ell$ as
\begin{equation}
\partial_t\bar{\rho}_\ell+\nabla\cdot\left(\sqrt{\bar{\rho}_\ell}\widehat{\bf w}_\ell\right)=0,
\label{qEuler-J-1-hat}
\end{equation}
\begin{equation}
\partial_t\left(\sqrt{\bar{\rho}_\ell}\widehat{\bf w}_\ell\right)+\nabla\cdot\overline{({\bf w}{\bf w})}_\ell=-\nabla \bar{p}_\ell+\nabla\cdot\bar{{\bf \Sigma}}_\ell+\bar{{\bf f}}_\ell.
\label{qEuler-J-2-hat}
\end{equation}

We now consider the large-scale kinetic energy balance.
We first note that, because $|{\bf v}|^2$ is a convex function of ${\bf v}$, the following inequality holds:
\begin{equation}
\dfrac{1}{2}|\widehat{\bf w}_\ell|^2=\dfrac{1}{2}\bar{\rho}_\ell\left|\dfrac{\overline{(\rho{\bf v})_\ell}}{\bar{\rho}_\ell}\right|^2\le\dfrac{1}{2}\bar{\rho}_\ell\dfrac{\overline{(\rho|{\bf v}|^2)}_\ell}{\bar{\rho}_\ell}=\dfrac{1}{2}\overline{(\rho|{\bf v}|^2)}_\ell.
\end{equation}
Therefore, the integral over space of $|\widehat{\bf w}_\ell|^2/2$ is less than the total kinetic energy:
\begin{align}
\int_\Omega d^d{\bf x}\dfrac{1}{2}|\widehat{\bf w}_\ell|^2\le\int_\Omega d^d{\bf x}\dfrac{1}{2}\rho|{\bf v}|^2,
\end{align}
and thus $|\widehat{\bf w}_\ell|^2/2$ represents the large-scale kinetic energy as in the one based on the Favre averaging \cite{Eyink_Drivas,tanogami2021theoretical}.
From (\ref{qEuler-J-1-hat}) and (\ref{qEuler-J-2-hat}), we obtain the large-scale kinetic energy balance:
\begin{align}
\partial_t\dfrac{1}{2}|\widehat{\bf w}_\ell|^2+\nabla\cdot{\bf J}_\ell&=\bar{p}_\ell\nabla\cdot\dfrac{\overline{\sqrt{\rho}}_\ell\bar{\bf w}_\ell}{\bar{\rho}_\ell}-\bar{\bf \Sigma}_\ell:\nabla\dfrac{\overline{\sqrt{\rho}}_\ell\bar{\bf w}_\ell}{\bar{\rho}_\ell}+\epsilon^{\mathrm{in}}_\ell\notag\\
&\quad-Q^{\mathrm{flux}}_\ell-\nabla\dfrac{\widehat{\bf w}_\ell}{\sqrt{\bar{\rho}_\ell}}:\left(\widehat{\bf w}_\ell\widehat{\bf w}_\ell-\bar{\bf w}_\ell\bar{\bf w}_\ell\right).
\label{large scale energy balance}
\end{align}
Here, $\epsilon^{\mathrm{in}}_\ell:=\widehat{\bf w}_\ell\cdot\bar{\bf f}_\ell/\sqrt{\bar{\rho}_\ell}$ denotes the energy injection rate due to external stirring at the scale $\ell$ and ${\bf J}_\ell$ represents the spatial transport of large-scale kinetic energy:
\begin{align}
{\bf J}_\ell&:=\dfrac{1}{2}|\widehat{\bf w}_\ell|^2\dfrac{\widehat{\bf w}_\ell}{\sqrt{\bar{\rho}_\ell}}+\bar{p}_\ell\dfrac{\overline{\sqrt{\rho}}_\ell\bar{\bf w}_\ell}{\bar{\rho}_\ell}-\bar{\bf \Sigma}_\ell\cdot\dfrac{\overline{\sqrt{\rho}}_\ell\bar{\bf w}_\ell}{\bar{\rho}_\ell}\notag\\
&\quad+\dfrac{\widehat{\bf w}_\ell}{\sqrt{\bar{\rho}_\ell}}\cdot\bar{\tau}_\ell({\bf w},{\bf w})-\dfrac{\widehat{\bf w}_\ell}{\sqrt{\bar{\rho}_\ell}}\cdot\left(\widehat{\bf w}_\ell\widehat{\bf w}_\ell-\bar{\bf w}_\ell\bar{\bf w}_\ell\right),
\end{align}
where $\bar{\tau}_\ell(f,g):=\overline{(fg)}_\ell-\bar{f}_\ell\bar{g}_\ell$.
The first two terms on the right-hand side of (\ref{large scale energy balance}) are the large-scale pressure-dilatation and quantum-stress--strain terms.
Because these two terms contain no modes at scales $<\ell$, they contribute only to the conversion of the large-scale kinetic energy into the interaction or quantum energies, and vice versa.
The last term on the right-hand side of (\ref{large scale energy balance}), which does not exist in Eq.\ (36) of \cite{tanogami2021theoretical}, arises from introducing the density-weighted variable $\widehat{\bf w}_\ell$.
Because this term contains $\widehat{\bf w}_\ell\widehat{\bf w}_\ell-\bar{\bf w}_\ell\bar{\bf w}_\ell$, we expect its contribution to the energy balance to be small relative to the other terms.
The term $Q^{\mathrm{flux}}_\ell$ represents the scale-to-scale kinetic energy flux:
\begin{equation}
Q^{\mathrm{flux}}_\ell:=\Pi_\ell+\Lambda^{(p)}_\ell+\Lambda^{(\Sigma)}_\ell.
\end{equation}
Here, $\Pi_\ell$ is the deformation work,
\begin{equation}
\Pi_\ell:=-\nabla\dfrac{\widehat{\bf w}_\ell}{\sqrt{\bar{\rho}_\ell}}:\bar{\tau}_\ell({\bf w},{\bf w}),
\end{equation}
$\Lambda^{(p)}_\ell$ is the baropycnal work,
\begin{equation}
\Lambda^{(p)}_\ell:=\dfrac{1}{\bar{\rho}_\ell}\nabla\bar{p}_\ell\cdot\bar{\tau}_\ell(\sqrt{\rho},{\bf w}),
\end{equation}
and $\Lambda^{(\Sigma)}_\ell$ is the quantum baropycnal work,
\begin{equation}
\Lambda^{(\Sigma)}_\ell:=-\dfrac{1}{\bar{\rho}_\ell}\nabla\cdot\bar{\bf \Sigma}_\ell\cdot\bar{\tau}_\ell(\sqrt{\rho},{\bf w}).
\end{equation}
Note that only these three terms are capable of direct transfer of kinetic energy across scales because each of the three terms has the form ``large-scale ($>\ell$) quantity $\times$ small-scale ($<\ell$) quantity,'' whereas the other terms on the right-hand side of (\ref{large scale energy balance}) do not.
We remark that these energy fluxes are not Galilean invariant, unlike those defined in \cite{tanogami2021theoretical}.
Therefore, they can be scale-local only if spatially or ensemble averaged \cite{Aluie_Eyink_2009-2}.

Let $\delta{\bf a}({\bf r};{\bf x}):={\bf a}({\bf x}+{\bf r})-{\bf a}({\bf x})$ for any field ${\bf a}$.
Instead of assuming the Besov regularity of the velocity field ${\bf v}$, we here assume the Besov regularity of the density-weighted velocity ${\bf w}$:
\begin{align}
\|\delta{\bf w}({\bf r};\cdot)\|_p\sim A_pw_0\left(\dfrac{|{\bf r}|}{L}\right)^{\sigma_p}\quad\text{as}\quad |{\bf r}|/L\rightarrow0
\label{Besov w}
\end{align}
with a dimensionless constant $A_p$ for $p\in[1,\infty]$ and $\sigma_p\in(0,1]$, where $w_0$ denotes the typical density-weighted velocity.
The important point here is that the validity of the Besov regularity (\ref{Besov w}) has been partially verified in several numerical simulations \cite{Kobayashi-Ueda,krstulovic2016grid,Leoni,Shukla_etal_2019}.
In addition, we assume the following properties for $\sqrt{\rho}$ instead of $\rho$:
\begin{align}
\|\delta\sqrt{\rho}({\bf r};\cdot)\|_p&=O(|{\bf r}|/L)\quad\text{as}\quad |{\bf r}|/L\rightarrow0,
\end{align}
\begin{align}
\|1/\sqrt{\bar{\rho}_\ell}\|_\infty&<\infty\quad\text{for}\quad\ell\ge\xi.
\end{align}
Under these assumptions, we can derive the same results as in \cite{tanogami2021theoretical}: 
\begin{align}
\|\Pi_\ell\|_{p/3}&=O\left(\left(\dfrac{\ell}{L}\right)^{3\sigma_p-1}\right),\notag\\
\|\Lambda^{(p)}_\ell\|_{p/3}&=O\left(\left(\dfrac{\ell}{L}\right)^{\sigma_p+1}\right),\notag\\
\|\Lambda^{(\Sigma)}_\ell\|_{p/3}&=O\left(\left(\dfrac{\ell}{L}\right)^{\sigma_p-1}\right),
\end{align}
for $p\ge 3$.
Following the same argument as in \cite{tanogami2021theoretical}, we thus find that the asymptotic behavior of the energy spectrum reads
\begin{align}
E(k)\sim
\begin{cases}
C_{\mathrm{large}}k^{-5/3}\quad\text{for}\quad\ell^{-1}_{\mathrm{large}}\ll k\ll\ell^{-1}_i,\\
C_{\mathrm{small}}k^{-3}\quad\text{for}\quad\ell^{-1}_i\ll k\ll\ell^{-1}_{\mathrm{small}},
\end{cases}
\end{align}
where $C_{\mathrm{large}}$ and $C_{\mathrm{small}}$ are positive constants. 
Note that $E(k)$ denotes the standard energy spectrum, not the velocity power spectrum $E^v(k)$ used in \cite{tanogami2021theoretical}.
We also remark that the $k^{-3}$ spectrum for the quantum stress cascade can become shallower because of the Kelvin-wave oscillations, as already mentioned in \cite{tanogami2021theoretical}.
The important point here is that the quantum stress cascade, induced by the quantum baropycnal work $\Lambda^{(\Sigma)}_\ell$, emerges following the Richardson cascade.
This is nothing but the main claim of our paper \cite{tanogami2021theoretical}.

\bibliography{Theoretical_analysis_of_quantum_turbulence_using_the_Onsager_ideal_turbulence_theory}

\end{document}